\documentstyle[12pt]{article}
\begin{document}

\title{Generally relativistical Daffin-Kemmer formalism
       and behaviour of quantum-mechanical particle of spin $1$
       in the Abelian monopole field}

\author {V.M.Red'kov \\
     Institute of Physics, Belarus Academy of Sciences\\
F Skoryna  Avenue 68, Minsk 72, Republic ob Belarus\\
e-mail: redkov@dragon.bas-net.by}

\maketitle

\begin{abstract}

It is shown that the~manner of introducing  the~interaction
between a~spin  $1$  particle  and  external  classical
gravitational field can be successfully unified  with~the approach  that
occurred with regard to a~spin $1/2$ particle and was first  developed
by  Tetrode,  Weyl,  Fock,  Ivanenko. On  that way a generally relativistical
Duffin-Kemmer equation  is costructed.   So,
the manner of extending the~flat  space  Dirac  equation  to
general relativity case indicates clearly that the~Lorentz  group
underlies equally both these theories. In other words, the~Lorentz
group retains its importance and significance  at  changing  the
Minkowski space  model  to  an~arbitrary  curved  space-time.  In
contrast to this,  at  generalizing  the~Proca   formulation,  we
automatically destroy any relations to the~Lorentz group, although
the~definition itself for a~spin $1$ particle as an~elementary object
was  based on  just  this  group.  Such  a~gravity's sensitiveness to
the~fermion-boson division might appear rather  strange  and  unattractive
asymmetry, being subjected to the~criticism. Moreover,  just  this
feature  has brought about a~plenty  of  speculation  about  this
matter. In any case, this peculiarity of particle-gravity  field
interaction  is  recorded  almost  in  every  handbook.

In the paper,  on the base  of  the Duffin-Kemmer formalism developed,
the problem of a vector particle in the~Abelian  monopole
potential is considered.
The content is:
1. On Duffin-Kemmer formalism in the Rimannian space-time;
2. On wave functions of a spin 1 particle in the monopole field;
3. On connection with the Proca approach; 4. Discret symmetry.

\end{abstract}

\newpage

\subsection*{1.   On Duffin-Kemmer formalism
                 in the Rimannian space-time}

A generally acceptable point of view is that   description
of interaction between a quantum mechanical particle and an~external
classical  gravitational  field  looks  basically  different   in
accordance as whether   fermion  or  boson  is  meant.  So,  the
starting flat space (Dirac) equation
$$
(\; i \gamma ^{a} \;  \partial _{a} \; - \; m \;) \; \Psi (x) \; = \; 0
$$

\noindent as well known, we have to generalize through the~use of the~tetrad
formalism  according  to  the~Tetrode-Weyl-Fock-Ivanenko   (TWFI)
procedure  [1-9].  With regard to a vector bosons [10-33],  a~totally
different approach is generally  used:  it  consists  in  ordinary
formal changing all involved tensors and usual  derivative
$\partial_{a}$
into  general relativity ones. For example, in  case  of
a~vector (spin 1) particle, the~flat space Proca equations
$$
( \partial _{a} \; \Psi _{b} \; - \;
\partial _{b} \; \Psi _{a}) = \;m \qquad  \Psi _{ab} \;\; , \;\;
\partial ^{b} \; \Psi _{ab} = m \;  \Psi  _{a}
\eqno(1.1a)
$$

\noindent being subjected to the~formal change
$$
\partial _{a} \; \rightarrow \;  \nabla _{\alpha } \; , \qquad
\Psi _{a} \; \rightarrow \; \Psi _{\alpha } ,\qquad
\Psi _{ab} \;  \rightarrow  \; \Psi _{\alpha \beta }
\eqno(1.1b)
$$

\noindent results in
$$
(\nabla _{\alpha }\; \Psi _{\beta} - \nabla _{\beta }\; \Psi_{\alpha }) =
m \; \Psi _{\alpha \beta } \; , \qquad
\nabla ^{\beta }\; \Psi _{\alpha \beta } = m \; \Psi _{\alpha }
\eqno(1.1c)
$$

\noindent However, it is  known  already  for  a~long  time  that  all
particles of the~theory, irrespective of whether bosons or fermions
are meant, obey  in a curved background space-time a~unique
TWFI approach  (see,  for  example,  in  [8,9]).  But  admittedly,
in the~common literature,  they   do  not  use  consistently  this
universal formalism. Although the~widely spread  method  of  light
tetrad  or Newman-Penrose formalism  [34,35]) is certainly a~renewed and modified
variant  of  the  TWFI   above mentioned approach,  the~Newman-Penrose method
was  developed   in  accordance  with  its  own   special
intrinsic requirements and with no clearly  visible  relations  to
the conventional TWFI approach (such a~correlation is  potentially
implied rather than observed really).

As a matter of fact, a potentially  existing  (general  relativity)
Duffin-Kemmmer $(D-K)$ equation for a spin 1  particle,  apparently,
is not  widely  adopted.  But,  as  evidenced  by  many  examples,
sometimes it is desirable  if  not  necessary,  to  depart  from
constructions of common use in order to arrive  at  a~simpler  or
more suitable one for a~particular situation. Bellow,  we  develop
some aspects of this generalized $D-K$ theory, that are essential
to real practical calculations (I adhere an unpublished work
of the three authors [...]).  This  method
will be  successfully  applied  further  in  Sec.2  to  a~spin  $1$
particle-monopole problem.

So, let us take up considering this matter in more detail. We
start from a flat space equation  in  its  matrix  (Duffin-Kemmer)
form [10]:
$$
(\; i\;  \beta ^{a} \; \partial_{a} \; - \; m \; )\;  \Phi  (x) = 0
\eqno(1.2a)
$$

\noindent where $\Phi (x)$ is  a ten component  column-function;
$\beta ^{a}$  is $(10 \times 10)$ -matrices; in the Cartesian representation
they are
$$
\Phi  = ( \Phi _{0} \; \Phi _{1} \; \Phi _{2} \Phi _{3} ; \;
\Phi _{01} \; \Phi _{02} \; \Phi _{03} ; \; \Phi _{23}\; \Phi _{31} \;
\Phi _{12} )                     \;\;  ,
$$
$$
\beta ^{a} = \pmatrix{0 & \kappa ^{a} \cr \lambda ^{a} & 0} =
 ( \kappa  ^{a} \oplus \lambda  ^{a} ) \;, \;\;
( \kappa  ^{a})_{j} ^{[mn]} \; = \;
- i ( \delta ^{m}_{j} \; g^{na} \;  - \; \delta  ^{n}_{j} \; g^{ma} )\;\; ,
$$
$$
( \lambda  ^{a})^{j}_{[mn]} \; = \;
  - i ( \delta  ^{a}_{m} \; \delta ^{j}_{n}  -
\delta ^{a}_{n} \;\delta ^{j}_{m} ) \; = - i \delta ^{aj}_{mn}
\eqno(1.2b)
$$

\noindent $( g^{na} ) = \hbox{diag}( +1,-1,-1,-1 )$  is the  Minkowski
 metric  tensor; the sectional  matrix  structure  introduced  here
 will  be  used bellow.  By using this representation (5.2b),
we can easily verify the major properties of $\beta ^{a}$:
$$
\beta ^{c} \; \beta ^{a} \; \beta ^{b} =
\left( \begin{array}{cc}
     0 & \kappa ^{c} \;\lambda ^{a}\;\kappa ^{b} \\
\lambda^{c} \; \kappa ^{a} \; \lambda ^{b} & 0
\end{array} \right ) ,  \qquad
(\lambda ^{c} \; \kappa ^{a} \; \lambda ^{b}) ^{j}_{[mn]} =
 i\; [ \delta ^{cb}_{mn} \; g^{aj} \;  - \; \delta ^{cj}_{mn} \;  g^{ab} ] \; ,
$$
$$
(\kappa ^{c} \; \lambda ^{a} \; \kappa ^{b})^{[mn]}_{j}  =
 i \; [ \delta ^{a}_{j}\; (g^{cm} \; g^{bn} \; - \; g^{cn}\; g^{bm} ) \; + \;
\;g^{ac}\; ( \delta ^{n}_{j} \; g^{mb} \;  - \; \delta ^{m}_{j}\; g^{nb} ) ]
$$

\noindent and then
$$
( \beta ^{c} \; \beta ^{a} \; \beta  ^{b} \;  + \;
\beta  ^{b} \; \beta ^{a} \; \beta ^{c} ) = ( \beta ^{c} \; g^{ab} \; + \;
\beta ^{b} g^{ac} ) ,
$$
$$
[\beta ^{c} , j^{ab} ] = ( g^{ca} \;\beta^{b} \; - \;
g^{cb} \; \beta ^{a} )  , \qquad
j^{ab} = ( \beta ^{a} \; \beta ^{b} \; - \; \beta ^{b} \; \beta ^{a} ) ,
$$
$$
[j^{mn}, j^{ab}]  = ( g^{na} \; j^{mb} \;  - \; g^{nb} \; j^{ma} ) \; - \;
( g^{ma} \;j^{nb}\;   - \; g^{mb} \; j^{na} ))
$$

To follow the TWFI procedure,  the equation (1.2a)  must  be
extended to a Rimannian space-time (with a metric $g_{\alpha \beta }(x)$
and its concomitant tetrad $e^{\alpha }_{(a)}(x) )$ according to
$$
[ \; i \; \beta ^{\alpha }(x)\; (\partial_{\alpha} \;  +  \;
B_{\alpha }(x) ) \; - m \;  ] \;\Phi  (x)  = 0
\eqno(1.3)
$$

\noindent where
$$
\beta ^{\alpha }(x) = \beta ^{a} e ^{\alpha }_{(a)}(x) \; , \qquad
B_{\alpha }(x) =
{1 \over 2}\; j^{ab} e ^{\beta }_{(a)}\nabla _{\alpha }( e_{(b)\beta }) , \qquad
j^{ab} = ( \beta ^{a} \beta ^{b} - \beta ^{b} \beta ^{a})  .
$$

\noindent
This equation  contains  the  tetrad $e^{\alpha }_{(a)}(x)$
explicitly. Therefore, there must  exist  a~possibility  to  demonstrate
 the~equivalence of  any  variants  of  this  equation
associated with various tetrads:
$$
e^{\alpha }_{(a)}(x)  \;\; \hbox{and} \;\;  e'^{\alpha }_{(b)}(x)\; = \;
 L^{b}_{a} (x) \; e^{\alpha }_{(b)}(x)
\eqno(1.4a)
$$

\noindent $(L(x)$ is an arbitrary local Lorentz transformation).
We will  show  that  two such equations
$$
[\; i\; \beta ^{\alpha }(x) \; (\partial_{\alpha} \; + \;
B_{\alpha }(x)) \;  -  \; m ] \; \Phi  (x)  = 0 \; , \qquad
[\;i\; \beta'^{\alpha }(x)\; (\partial_{\alpha}  \;+\; B'_{\alpha }(x))
 \; - \; m\; ] \; \Phi'(x)  = 0
\eqno(1.4b)
$$

\noindent generating in tetrads $e^{\alpha }_{(a)}(x)$    and
 $e'^{\alpha }_{(b)}(x)$,
respectively,  can  be  converted  into  each  other  through  the
transformation $\Phi (x) = S(x)\; \Phi (x)$ :
$$
\left ( \begin{array}{c}
                   \phi'_{a}(x) \\ \phi'_{[ab]}(x)
\end{array} \right ) =
\left ( \begin{array}{cc}
         L_{a}^{\;\;l} & 0 \\
         0 &  L_{a}^{\;\;m} L_{b}^{\;\;n}
\end{array} \right ) \;\;
\left ( \begin{array}{c}
                   \phi_{l}(x) \\ \phi_{[mn]}(x)
\end{array} \right )
\eqno(1.4c)
$$

\noindent here the $L(x)$ is the same as in the~relation (1.4a).
So, starting from the first equation in (1.4b), let us obtain
an equation for $\Phi'(x)$. Allowing for $\Phi(x) = S(x)\; \Phi (x)$, we get
$$
[\; i\; S \; \beta ^{\alpha } \; S^{-1} (\partial _{\alpha} \; + \;
S \; B_{\alpha }\; S^{-1} \; + \; S \; \partial_{\alpha}\; S^{-1}) \; -
 \; m \;] \; \Phi'(x) = 0
$$

\noindent A task that faces us now is of verifying the relationships
$$
S(x) \; \beta ^{\alpha }(x) \; S^{-1}(x) =
\beta'^{\alpha}(x)      \;\; ,
\eqno(1.5a)
$$
$$
[ S(x) \; B(x)\; S^{-1}(x) \; + \; S(x) \; \partial_{\alpha}\; S^{-1}(x)\; ]
 = 0       \; .
\eqno(1.5b)
$$

\noindent The first one can be rewritten as
$$
S(x) \; \beta ^{a} \; e ^{\alpha }_{(a)}(x) \; S^{-1}(x) \; =
\beta ^{b} \; e'^{\alpha }_{(b)}(x)
$$

\noindent from where,  taking  into  account  the  relation (1.4a)  between
tetrads, we come to
$$
S(x) \;\beta ^{a} \; S^{-1}(x) \; = \; \beta ^{b} \; L^{a}_{b}(x) \; .
\eqno(1.5c)
$$

\noindent The latter condition is of great familiarity in $D-K$  theory;  one
can verify it through the use of the sectional  structure  of $\beta ^{a}$,
which provides two sub-relations:
$$
L(x) \; \kappa ^{a} \; [\; L^{-1}(x) \otimes  L(x)^{-1}\;] \; =
 \;\kappa ^{b} \; L^{\;\;a}_{b}(x) \; , \qquad
[\; L(x) \otimes  L(x)\; ] \; \lambda ^{a} \; L(x)^{-1} \; = \;
\lambda ^{b} \; L^{\;\;a}_{b}(x) \; .
\eqno(1.5d)
$$

\noindent Those latter will be satisfied identically, after we take explicit
form of $\kappa ^{a}$ and $\lambda ^{a}$  into account and  also allow
for  the $L^{\;\;b}_{a}$
being pseudo orthogonal: $g^{al} \; (L^{-1})^{\;\;k}_{l}(x) =
g^{kb}\; L^{\;\;a}_{b}(x)$.
Now, let us pass to the proof of the relationship (1.5b).  By
using the determining relation for $D-K$  connection
$$
B_{\alpha }(x) = {1 \over 2} \; j ^{ab} \; e^{\beta }_{(a)} \;
 \nabla _{\alpha } \; ( e_{(b)\beta }) , \qquad j^{ab} \; =
(\beta ^{a} \; \beta ^{b} \; - \; \beta ^{b} \; \beta ^{a})
$$

\noindent and also the formula (1.5c), we get
$$
S(x)\; \partial_{\alpha} \; S^{-1}(x) \; = \;  [ \; B'_{\alpha }(x) \; - \;
{1 \over 2} \; j^{mn} L^{\;\;n}_{m}(x)\; g_{ab} \; \partial _{\alpha}\;
 L^{\;\;b}_{n}(x) \;] \; ,
$$

\noindent In a sequence, the (1.5b) results in
$$
S(x) \; \partial_{\alpha} \; S^{-1}(x) =  \;
{1\over2}\;  L^{\;\;a}_{m}(x)\; g_{ab}\;
(\; \partial_{\alpha } \;L^{\;\;b}_{n}(x)\; ) \; .
$$

\noindent The latter condition is an identity: this is  readily verified
through the~use of sectional structure of all involved matrices.

Thus, the equations from (1.4b) are  translated  into  each
other; thereby,  they  manifest  a~gauge  symmetry under
local Lorentz transformations (in a~complete
analogy with more familiar Dirac particle case  [1-9]).  In  the~same
time, the~wave function  from  this  equation  represents  scalar quantity
relative to general coordinate transformations: that is,
if $\; x^{\alpha } \; \rightarrow  \; x'^{\alpha } =
f^{\alpha }(x)$ ,  then $\; \Phi'(x) = \Phi (x)$.

It remains to demonstrate that this $D-K$  formulation  can  be
inverted into the Proca formalism in terms of  general  relativity
tensors. To this end, as a~first  step,  let  us  allow  for
the~sectional structure of $\beta ^{a}, J^{ab}$  and $\Phi (x)$ in
the~$D-K$  equation; then instead of (1.3) we get
$$
i \; [\; \lambda^{c} \; e^{\alpha }_{(c)} \; (\; \partial_{\alpha} \;+ \;
\kappa ^{a} \; \lambda ^{b} \; e^{\beta }_{(a)} \;
\nabla_{\alpha }\; e_{(b)\beta }\; ) \; ]^{\;\;\;\;\;\;l}_{[mn]}\; \Phi _{l} =
m\; \Phi _{[mn]}    \; ,
$$
$$
i \; [\kappa ^{c} \; e^{\alpha }_{(c)} \; ( \;\partial_{\alpha} \; + \;
\lambda ^{a} \; \kappa ^{b}\; e^{\beta }_{(a)} \;
\nabla _{\alpha } \; e_{(b)\beta }\; ) \; ]^{[\;\;mn]}_{l} \; \Phi_{[mn]}  =
m \Phi _{l}
\eqno(1.6a)
$$

\noindent which, after taking into account the~explicit form of
 $( \lambda ^{c},\;  \lambda ^{c} \; \kappa ^{a} \; \lambda ^{b}$  ,
$\kappa ^{c} ,\;  \kappa ^{c} \; \lambda ^{a} \; \kappa^{b} )$, lead to
$$
[\; (e_{(a)}^{\alpha} \; \partial_{\alpha} \; \Phi _{b} \; - \;
e^{\alpha }_{(b)} \; \partial_{\alpha}\; \Phi _{a}) \; + \;
( \gamma ^{c}_{\;\;ab} - \gamma ^{c}_{\;\;ba} ) \; \Phi _{c}\; ] = m \; \Phi _{ab}\; ,
$$
$$
[\; e^{(b)\alpha } \;\partial_{\alpha } \; \Phi _{ab} \; + \;
\gamma ^{nb} _{\;\;\;\;n} \Phi _{ab} + \gamma ^{\;\;mn}_{a} \Phi _{mn}\; ]  =
m \; \Phi _{a}  \; .
\eqno(1.6b)
$$

\noindent In turn, these will represent just  the~Proca equations (1.1c)
after they are rewritten in terms of tetrad  components  according  to
$$
\Phi _{a}(x) \;= \; e^{\alpha }_{(a)} (x) \; \Phi _{\alpha }(x) , \qquad
\Phi _{ab}(x)= e^{\alpha }_{(a)} (x) \;  e^{\beta }_{(b)}(x) \;
\Phi_{\alpha \beta }(x)
\eqno(1.7)
$$

\noindent the  symbol $\gamma _{abc}(x)$ is used to designate  a~rotational
Ricci  coefficients:
$$
\gamma _{abc}(x) \; =  \;
- \; (\nabla _{\beta } \;
e_{(a)\alpha }) \; e^{\alpha }_{(b)} \; e^{\beta }_{(c)} \; .
$$

So, as evidenced by the~above, the~manner of introducing  the~interaction
between a~spin  $1$  particle  and  external  classical
gravitational field can be successfully unified  with~the approach  that
occurred with regard to a~spin $1/2$ particle and was first  developed
by  Tetrode,  Weyl,  Fock,  Ivanenko.  One  should   attach   great
significance to that possibility of unification.  Moreover,  its
absence would  be a~very strange fact indeed  because  it  touches
concepts of great  physical  significance.  Let  us  discuss  this
matter in more detail.

The manner of extending the~flat  space  Dirac  equation  to
general relativity case indicates clearly that the~Lorentz  group
underlies equally both these theories. In other words, the~Lorentz
group retains its importance and significance  at  changing  the
Minkowski space  model  to  an~arbitrary  curved  space-time.  In
contrast to this,  at  generalizing  the~Proca   formulation,  we
automatically destroy any relations to the~Lorentz group, although
the~definition itself for a~spin $1$ particle as an~elementary object
was  based on  just  this  group.  Such  a~gravity  sensitiveness to
the~fermion-boson division might appear rather  strange  and  unattractive
asymmetry, being subjected to the~criticism. Moreover,  just  this
feature  has brought about a~plenty  of  speculation  about  this
matter. In any case, this peculiarity of particle-gravity  field
interaction  is  recorded  almost  in  every  handbook.
By my mind, the~possibility  itself  of  rewriting
the~tetrad-based Duffin-Kemmer equation in terms of general  relativity
tensors  looks very  surprising  indeed.

\subsection*{2.  On wave functions of a spin $1$ particle
in the monopole field}

\vspace{10mm}

Now, on the base  of  Duffin-Kemmer (D-K)  formalism,  let  us
consider the problem of a vector particle in the~Abelian  monopole
potential. The starting D-K equation in the~spherical tetrad takes
the~form
$$
\left [\;i\;\beta ^{0} \; \partial _{t} \; +
\; i \; ( \beta ^{3} \; \partial _{r}\; +
{1 \over r} \;(\beta ^{1} \; j^{31}  + \beta ^{2} \; j^{32}) ) \;+
\; {1 \over r} \;  \Sigma ^{\kappa }_{\theta ,\phi } \; -
\; {mc \over \hbar} \;\right ]
\;  \Phi (x)  = 0
\eqno(2.1a)
$$

\noindent where
$$
\Sigma ^{\kappa }_{\theta ,\phi } \; = \;
\left [\; i\; \beta ^{1} \; \partial _{\theta } \; + \;
\beta ^{2} \; {i\; \partial \; + \; (i\; j^{12} - \kappa )
 \cos \theta  \over \sin \theta} \right ]
\eqno(2.1b)
$$

\noindent These relations are very close  to  analogous  ones  used  in
 the~electronic case [36] ; variations concern  only  the~explicit  expressions
for matrices:
 $\gamma ^{a} , \; \sigma ^{ab} $ are to be changed into
$\; \beta ^{a} , \; J^{ab}$.

Below,  we  will  use  the~cyclic basis for Duffin-Kemmer matrices:

$$
\beta^{0} = \left (   \begin{array}{cccccccccc}
0  &  0   &   0   &  0     &  0  &  0  &   0      &  0   & 0   &  0  \\
0  &  0   &   0   &  0     & +i  &  0  &   0      &  0   & 0   &  0  \\
0  &  0   &   0   &  0     &  0  & +i  &   0      &  0   & 0   &  0  \\
0  &  0   &   0   &  0     &  0  &  0  &  +i      &  0   & 0   &  0  \\
0  & -i   &   0   &  0     &  0  &  0  &   0      &  0   & 0   &  0  \\
0  &  0   &  -i   &  0     &  0  &  0  &   0      &  0   & 0   &  0  \\
i  &  0   &   0   & -i     &  0  &  0  &   0      &  0   & 0   &  0  \\
0  &  0   &   0   &  0     &  0  &  0  &   0      &  0   & 0   &  0  \\
0  &  0   &   0   &  0     &  0  &  0  &   0      &  0   & 0   &  0  \\
0  &  0   &   0   &  0     &  0  &  0  &   0      &  0   & 0   &  0
\end{array} \right )  ,
$$
\vspace{5mm}
$$
\beta^{3} = \left (   \begin{array}{cccccccccc}
0  &  0   &   0   &  0     &  0  &  i  &   0      &  0   & 0   &  0  \\
0  &  0   &   0   &  0     &  0  &  0  &   0      & +1   & 0   &  0  \\
0  &  0   &   0   &  0     &  0  &  0  &   0      &  0   & 0   &  0  \\
0  &  0   &   0   &  0     &  0  &  0  &   0      &  0   & 0   & -1  \\
0  &  0   &   0   &  0     &  0  &  0  &   0      &  0   & 0   &  0  \\
0  &  0   &   0   &  0     &  0  &  0  &   0      &  0   & 0   &  0  \\
i  &  0   &   0   &  0     &  0  &  0  &   0      &  0   & 0   &  0  \\
0  & -1   &   0   &  0     &  0  &  0  &   0      &  0   & 0   &  0  \\
0  &  0   &   0   &  0     &  0  &  0  &   0      &  0   & 0   &  0  \\
0  &  0   &  +1   &  0     &  0  &  i  &   0      &  0   & 0   &  0
\end{array} \right )  ,
$$
\vspace{5mm}
$$
\beta^{1} = {1 \over \sqrt{2}} \;
\left (   \begin{array}{cccccccccc}
0  &  0   &   0   &  0     & -i  &  0  &  +i      &  0   &  0  &  0  \\
0  &  0   &   0   &  0     &  0  &  0  &   0      &  0   & +1  &  0  \\
0  &  0   &   0   &  0     &  0  &  0  &   0      & +1   &  0  & +1  \\
0  &  0   &   0   &  0     &  0  &  0  &   0      &  0   & +1  &  0  \\
-i &  0   &   0   &  0     &  0  &  0  &   0      &  0   & 0   &  0  \\
0  &  0   &   0   &  0     &  0  &  0  &   0      &  0   & 0   &  0  \\
+i &  0   &   0   &  0     &  0  &  0  &   0      &  0   & 0   &  0  \\
0  &  0   &  -1   &  0     &  0  &  0  &   0      &  0   & 0   &  0  \\
0  & -1   &   0   & -1     &  0  &  0  &   0      &  0   & 0   &  0  \\
0  &  0   &  -1   &  0     &  0  &  0  &   0      &  0   & 0   &  0
\end{array} \right )  ,
$$
\vspace{5mm}
$$
\beta^{2} = {1 \over \sqrt{2}} \;
\left (   \begin{array}{cccccccccc}
0  &  0   &   0   &  0     &  1  &  0  &   1      &  0   &  0  &  0  \\
0  &  0   &   0   &  0     &  0  &  0  &   0      &  0   & -i  &  0  \\
0  &  0   &   0   &  0     &  0  &  0  &   0      & +i   &  0  & -i  \\
0  &  0   &   0   &  0     &  0  &  0  &   0      &  0   & +i  &  0  \\
-1 &  0   &   0   &  0     &  0  &  0  &   0      &  0   & 0   &  0  \\
0  &  0   &   0   &  0     &  0  &  0  &   0      &  0   & 0   &  0  \\
-1 &  0   &   0   &  0     &  0  &  0  &   0      &  0   & 0   &  0  \\
0  &  0   &  +i   &  0     &  0  &  0  &   0      &  0   & 0   &  0  \\
0  & -i   &   0   & +i     &  0  &  0  &   0      &  0   & 0   &  0  \\
0  &  0   &  -i   &  0     &  0  &  0  &   0      &  0   & 0   &  0
\end{array} \right )  ,
$$

\noindent correspondingly, the matrix $ij^{12}$  has a~diagonal structure
$$
ij^{12} = \left ( \begin{array}{cccc}
         0   &   0    &   0     &     0  \\
         0   &  t_{3} &   0     &     0  \\
         0   &   0    &  t_{3}  &     0  \\
         0   &   0    &   0     &    t_{3}
\end{array} \right )  , \qquad
 t_{3} = \left ( \begin{array}{ccc}
             +1   &   0   &   0  \\
              0   &   0   &   0  \\
              0   &   0   &  -1
\end{array} \right )      \; .
$$

In the given tetrad representation, three components of the~total
conserved momentum are  (compare with [37,38])
$$
j^{\kappa }_{1} = \left [ \; l_{1} + { \cos \phi \over \sin \theta}
 \;(ij^{12} - \kappa) \; \right ] \; ,
\qquad
j^{\kappa }_{2} =\left  [\;l_{2} + { \sin \phi \over \sin \theta} \;
(ij^{12} - \kappa) \; \right ] \; ,
\qquad
j^{\kappa }_{3} = l_{3} \; .
\eqno(2.2a)
$$

\noindent Correspondingly, according to  the~general procedure [36],
the~particle's wave functions with fixed quantum number
$(\epsilon , j , m )$ are to be constructed as follows:
$$
\Phi _{\epsilon jm}(x)  = e^{-i\epsilon t} \;
[ f_{1}(r)\; D_{\kappa } ,\;  f_{2}(r)\; D_{\kappa -1} , \; f_{3}(r) \; D_{\kappa},
\; f_{4}(r) \; D_{\kappa +1},
$$
$$
 f_{5}(r) \; D_{\kappa -1}, \;  f_{6}(r) \; D_{\kappa } , \;
 f_{7}(r) \; D_{\kappa +1}, \;
f_{8}(r) \;D_{\kappa -1} , \; f_{9}(r) \;D_{\kappa } ,\; f_{10}(r)\; D_{\kappa +1} ]
\eqno(2.2b)
$$

\noindent here, $D_{\sigma } \equiv  D^{j}_{-m,\sigma } (\phi ,\theta ,0)$.

At finding $10$ radial equations for $f_{1},\ldots, f_{10}$,  we  are  to
use the~six recursive relations [39]
$$
\partial_{\theta } \; D_{\kappa -1} \; = \;
(a \; D_{\kappa-2} - c \; D_{\kappa } ), \;\;
{ -m-(\kappa -1) \cos \theta  \over \sin \theta } \; D_{\kappa -1} =
(-a \; D_{\kappa -2} - c \;  D_{\kappa }),
$$
$$
\partial _{\theta }\; D_{\kappa } \;  = \;
(c \; D_{\kappa-1} -  d \; D_{\kappa +1}), \;\;
{- m  - \kappa  \cos \theta \over \sin \theta } \; D_{\kappa } =
(-c \;  D_{\kappa-1} - d \;  D_{\kappa +1}),
$$
$$
\partial _{\theta } \; D_{\kappa +1} \; = \;
(d \; D_{\kappa } - b \;  D_{\kappa +2}), \;\;
{-m-(\kappa +1)\cos \theta \over \sin \theta } \;  D_{\kappa +1} \; = \;
(-d \;  D_{\kappa } - b \; D_{\kappa +2})
$$

\noindent where
$$
a = {1 \over 2}  \sqrt{(j + \kappa  -1)(j - \kappa  + 2)}  , \qquad
b = {1 \over 2}  \sqrt{(j - \kappa  -1)(j + \kappa  + 2)} ,
$$
$$
c = {1 \over 2}  \sqrt{(j + \kappa )(j - \kappa + 1)} , \qquad
d = {1 \over 2}  \sqrt{(j - \kappa )(j + \kappa  +1)} .
$$

Allowing for the following intermediate results
$$
\Sigma ^{\kappa }_{\theta ,\phi } \; \Phi\;  = \;
\exp{-i\epsilon t} \; \sqrt{2} \;
\left ( \begin{array}{c}
   (\; -\; c \; f_{5} \; -\;  d \; f_{7}\; )\;\;  D_{\kappa} \\
  -\; i\; c \; f_{9} \;\; D_{\kappa -1}            \\
  (\; -\; i\; c \; f_{8} \;+ \;i \; d \; f_{10}) \; \; D_{\kappa} \\
   -\; i \; d\; f_{9} \;\; D_{\kappa +1}           \\
       c \; f_{1} \;\; D_{\kappa -1}  \\  0   \\
       d \; f_{1} \;\; D_{\kappa +1}  \\
    - \;i\; c \; f_{3} \;\; D_{\kappa -1}  \\
  (\;+\; i\; c \; f_{2} \; - \; i\;  d \; f_{4})\;\;  D_{\kappa} \\
   +\; i\; d \; f_{3} \;\; D_{\kappa +1}
\end{array} \right ) ;
\eqno(2.3a)
$$
$$
i \; \beta ^{0} \; \partial _{t} \; \Phi  = \; \epsilon \; \exp{-i\epsilon t} \;
\left ( \begin{array}{c}
    0  \\  -\; i\; f_{5} \;\;  D_{\kappa - 1} \\  i\; f_{6}\; \; D_{\kappa }  \\
    i\; f_{7} \;\; D_{\kappa +1} \\  -\; i\; f_{2} \;\; D_{\kappa -1}  \\
   -\; i\;  f_{3} \;\; D_{\kappa}  \\     -\; i \; f_{4} \;\; D_{\kappa +1}  \\
     0 \\   0   \\  0
\end{array} \right )   \;  ;
\eqno(2.3b)
$$
$$
i \; (\; \beta ^{3} \; \partial _{r} \; + \; {1 \over r} \;
(\; \beta ^{1} \; \beta ^{31} + \beta ^{2} \; \beta ^{32})\; ) \;
 \Phi _{\epsilon jm}  = \; \exp{-i\epsilon t} \;
\left ( \begin{array}{c}
     (\; - \; d/dr \; - \; 2/r\; ) \; f_{6} \; \; D_{\kappa }  \\
     (\;i\;d/dr\; + \;i/r\;) \; f_{8} \;\; D_{\kappa -1} \\  0 \\
     (\;-\;i\;d/dr\; -\;i/r\;) \; f_{10}\;\; D_{\kappa +1} \\ 0 \\ 0 \\ 0 \\
     (\;-\;i\; d/dr \;-\; i/r\;) \; f_{2} \;\; D_{\kappa -1} \\ 0 \\
     (\; i\;d/dr \;+\;i/r\;) \; f_{4} \;\; D_{\kappa +1}
\end{array} \right )
\eqno(2.3c)
$$

\noindent from (2.1a) we produce
$$
-( {d \over dr} + {2 \over r} ) \; f_{6} - {\sqrt{2} \over r} \;
( c \; f_{5} + d \; f_{7})  - m \; f_{1} = 0       \; ,
$$
$$
i \epsilon \;  f_{5} + i ( {d \over dr} + {1 \over r} ) \; f_{8} +
i {\sqrt{2} c \over r}\; f_{9} - m \; f_{2} = 0      \; ,
$$
$$
i \epsilon \; f_{6} + {2i \over r} \; (- c \; f_{8} + d \; f_{10})
 - m \; f_{3} = 0                                        \; ,
$$
$$
i \epsilon \; f_{7} - i ( {d \over dr} + {1 \over r} ) \; f_{10}  - i {\sqrt{2} d \over r}\;
 f_{9}  - m \; f_{4}= 0                                      \; ,
$$
$$
i \epsilon \; f_{2} +  {\sqrt{2} c \over r} \; f_{1}  - m \; f_{5} = 0  , \;\;
-i \epsilon \; f_{3}  - {d \over dr} \; f_{1} - m \; f_{6} = 0   \; ,
$$
$$
-i \epsilon \; f_{4} +{\sqrt{2} d \over r} \; f_{1} - m \; f_{7} = 0 \; , \;\;
-i({d \over dr} + {1 \over r}) \; f_{2} - i {\sqrt{2} c \over r} \;  f_{3} - m \; f_{8} = 0
$$
$$
i{\sqrt{2} \over r} ( c \; f_{2}  - d \; f_{4})  - m \; f_{9} = 0 \; , \;\;
i ({d \over dr} + {1 \over r}) \; f_{4} + {i \sqrt{2} d \over r} \; f_{3}  - m \; f_{10} = 0\; .
\eqno(2.4)
$$

\noindent Parametre  $j$ are allowed to take  values  (we
have to draw distinction between $\kappa  = \pm  1/2$ and
all remaining $\kappa $):
$$
 \hbox{if} \;\;\; \kappa  = \pm  1/2 \;,
  \qquad \hbox{then} \;\;\; j = \mid \kappa \mid  ,
 \mid \kappa \mid  + 1 , \ldots ;
$$
$$
\hbox{if}\;\;\; \kappa = \pm  1, \pm  3/2,\ldots
\qquad  \hbox{then} \;\;\; j = \mid \kappa \mid - 1, \mid \kappa \mid ,
\mid \kappa \mid  + 1,\ldots
\eqno(2.5)
$$

\noindent In both cases, the states of minimal $j$ (respectively
 $j_{min.}= \mid \kappa \mid $  and
$j_{\min }= \mid \kappa \mid  - 1 )$  are to be considered separately:
the~radial system (2.4)  is not valid  for  those states.

Let us consider the state with $j_{\min } = \mid \kappa \mid -1$ .
 First,  one ought to investigate the $j_{min.} = 0$ situation arisen
at $\kappa  = \pm  1$;
the relevant wave function does not depend on the~$\theta ,\phi $  variables
at all.
Let $\kappa  = + 1$ and $j_{min.} = 0$, then we start
with  the~substitution
$$
\Phi ^{0}(t,r) = \; \exp{-i\epsilon t} \; [ \; 0 , \;\; \; f_{2} ,\;\;  0 , \;\;  0 ,
\;\; f_{5}, \;\; 0 ,\; 0 ; \;\; f_{8} , \;\; 0 , \;\; 0 \; ]
\eqno(2.6a)
$$

\noindent It is readily verified that the $\Sigma _{\theta ,\phi }$ operator
 acts on $\Phi _{0}$ as a~null  operator:
$\Sigma _{\theta ,\phi } \; \Phi _{0} \; = 0$;
because  the~identity $(i\; j^{12} \; - \; \kappa ) \; \Phi ^{0}\; \equiv\; 0$  holds.
 As a~result,  we produce only three non-trivial (as one should expect)
 equations:
$$
i \; \epsilon \; f_{5} \; + \; i \; (\;  {d \over dr}\; + \;{1 \over r}\; ) \;f_{8} - m \; f_{2}\; =  0
$$
$$
-\; i\; f_{2}\; -\; m \; f_{5} \;= 0 _ , \;\;\;  -\; i\;(\; {d \over dr}\; +\;
 {1 \over r})\;
 f_{2} - m \; f_{8} \; = 0
\eqno(2.6b)
$$

\noindent From here, it follows
$$
f_{5} = \; - \; i \; {\epsilon \over m} \; f_{2} , \qquad
f_{8}  = \; -\;  {i \over m} \;
( \; {d \over dr} \; + \; {1 \over r} ) \; f_{2}
$$

\noindent and the function $f_{2}$ ($F_{2} = {1 \over r}\; f_{2} $)  satisfies
the equation
$$
(\; {d^{2} \over dr^{2}} \; + \; \epsilon ^{2} \;   - \; m^{2} \; ) \; F_{2}  = 0
\eqno(2.6c)
$$

\noindent The latter provides us with an~exponential solution  of  the~same
kind as in the~electronic  case,  that  is  a~candidate   for
a~possible bound state.
The situation  with $j_{min.}= 0$   and $\kappa  = - 1$  looks
completely analogous:
$$
\Phi ^{0}(t,r)\; =\; \exp{-i\epsilon t} \; [\; 0, \;\; 0,\;\; 0,\;\; f_{4}\; ,\;\; 0,
\;\; 0, \;\; f_{7}, \;\;   0, \;\; 0, \;\; f_{10}\;  ]
\eqno(2.7a)
$$

\noindent and the radial equations
$$
i \; \epsilon\; f_{7}\; - i\;  ({d \over dr} \; + \; {1 \over r}) \; f_{10} \;-\; m \;f_{4} = 0 ,
$$
$$
-\; i \; f_{4} \; - \;m \; f_{7} = 0 , \qquad
 i\; ({d \over dr} \;+\; {1 \over r})\; f_{4} \; - m \; f_{10} = 0
\eqno(2.7b)
$$

\noindent and eventually we get

$$
f_{7} = -i\;{ \epsilon \over m} \; f_{4}  \;\; , \qquad
 f_{10}  = {i \over m } \; ({d \over dr} \;+ \;{1 \over r} ) \; f_{2} \;\; ,
$$
$$
(\; {d^{2} \over dr^{2}} \; + \; \epsilon ^{2} \; - \;m^{2} )\; F_{4} \;= 0 ,
\qquad (\; F_{4} = {1 \over r} \; f_{4}\; ) \; .
\eqno(2.7c)
$$

Now, we pass on the case  of  minimal $j_{min.} = \mid \kappa \mid -1$
with higher values of $\kappa $: $ \kappa = \pm 3/2 , \pm  2 , \ldots $
First,  let $\kappa $   be positive, then we have start with a~substitution
$$
\kappa  \ge  3/2 : \qquad
\Phi ^{0} \;=\; \exp{-i\epsilon t}\;  [ \;0 ,\; f_{2}\; D_{\kappa -1} ,\; 0 ,\; 0 ;
f_{5} \; D_{\kappa -1} ,\; 0 , \; 0 ;\; f_{8}\; D_{\kappa -1} ,\; 0 ,\; 0\; ]
\eqno(2.8a)
$$

\noindent Using the recursive relations
$$
\partial _{\theta}\; D_{\kappa-1} =  \sqrt{{\kappa -1 \over 2}}\;
 D_{\kappa -2} , \;\;
{-m - (\kappa -1) \cos \theta \over \sin \theta} \; D_{\kappa -1} =
 -  \sqrt{{\kappa -1 \over 2}}\; D_{\kappa -2}
$$
\noindent we find
$$
i \beta ^{1} \; \Phi ^{0}  = \exp{-i\epsilon t} \; i\;
\sqrt{{\kappa -1 \over 2}}
\left ( \begin{array}{c}
             -i f_{5} \; D_{\kappa -2} \\
              0  \\  + f_{8} \; D_{\kappa -2} \\  0  \\ 0  \\ 0  \\ 0\\ 0 \\
             -  f_{2} \;D_{\kappa -2} \\  0
\end{array} \right )
$$
\vspace{10mm}
$$
\beta ^{2}\; { i\; \partial _{\phi } + (ij^{12} - \kappa )\cos \theta  \over
 \sin \theta } \; \Phi ^{0} = e^{-i \epsilon t} \;
\sqrt{{\kappa -1 \over 2}} \;
\left ( \begin{array}{c}
   - f_{5} \; D_{\kappa - 2} \\  0 \\ -i f_{8}\; D_{\kappa-2}\\
     0 \\ 0 \\ 0 \\ 0 \\ 0  \\ +i f_{2} \; D_{\kappa-2} \\ 0
\end{array} \right )
$$

\noindent and further we produce $\Sigma _{\theta ,\phi } \Phi ^{0}  = 0$.
Therefore, the~radial functions $f_{2}, f_{5}, f_{8}$  satisfy  again
the~same  system (2.6b).
The  case  of $j_{min.}=\mid \kappa \mid -1$  with  negative $\kappa $  looks
completely similar to the above:
$$
\kappa  \leq -3/2: \qquad  \Phi ^{0} = e^{-i\epsilon t}\;
 [\; 0 ,\; 0 ,\; 0 ,\;\; f_{4}\; D_{\kappa +1} ,\; \; 0 ,\; 0 ,
$$
$$
\;\; f_{7}\; D_{\kappa +1} ,\; 0 ,\; 0 ,\;\; f_{10}\; D_{\kappa +1}\;\; ]
\eqno(2.9)
$$

\noindent the identity $\Sigma _{\theta ,\phi } \Phi ^{0} \equiv 0$ also
 holds  and  a~radial  system  coincides with (2.7b).
So, the  description  of $j_{min.} = \mid \kappa \mid - 1$  states has been
completed; all of them provide us  with  solutions  of  a~special
exponential kind which potentially might be related to a~bound state
and therefore  these solutions are of special physical  interest.  In
the~same time, unfortunately, it is a~unique case that we have managed  to  solve
entirely up to their radial equations.

Now, let us pass on the states with $j = \mid \kappa \mid $ that which are
to be regarded whether  as $j_{\min }= \mid \kappa \mid $  states  at
$\kappa  = \pm  1/2$  or non-minimal $j$ states at all other values of
 $\kappa $.
Let $j = \mid \kappa \mid $  and $\kappa $  be positive
 ($ \kappa  \geq  + 1/2$), then we have
to begin  with  a~substitution  (the~radial  functions  at  all
$D^{j = \kappa }_{-m,\kappa +1}$   in $\Phi (x)$ are equated to zero)

$
\kappa \geq  + 1/2 :$
$$
\Phi _{\epsilon jm}(x) \; = \; \exp{-i\epsilon t} \; [ \; f_{1}(r)\;D_{\kappa } ;\;
f_{2}(r)\; D_{\kappa -1} , \; f_{3}(r) \; D_{\kappa } ,\; 0 ;
$$
$$
f_{5}(r) \; D_{\kappa -1} ,\; f_{6}(r) \; D_{\kappa } , \; 0 ;
\; f_{8}(r)\;D _{\kappa -1} ,\; f_{9}(r) \; D_{\kappa } , \; 0 \; ]
\eqno(2.10a)
$$

\noindent For $\Sigma _{\theta ,\phi }\Phi $  we get

$$
\Sigma _{\theta ,\phi } \; \Phi =  \; \exp{-i\epsilon t} \sqrt{\kappa}\;
\left ( \begin{array}{c}
            - f_{5} \; D_{\kappa }  \\
           +i f_{9} \; D_{\kappa -1} \\
           -i f_{8} \; D_{\kappa }   \\
            0 \\ f_{1} \; D_{\kappa -1}\\  0 \\  0 \\
           -i f_{3} \; D_{\kappa -1} \\
           +i f_{2} \; D_{\kappa }  \\ 0
\end{array} \right )
$$

\noindent and further we produce the~radial system
$$
- ( {d \over dr} + {2 \over r} ) \; f_{6} - {\sqrt{\kappa} \over r} f_{5}
 - m \; f_{1} = 0 , \;\;
i \epsilon  \; f_{5} + i ( {d \over dr} + {1 \over r} )\; f_{8} +
i {\sqrt{\kappa} \over r} \; f_{9}  - m \; f_{2}  = 0  ,
$$
$$
i \epsilon \; f_{6} - i { \sqrt{\kappa} \over r} \; f _{8} - m \; f _{3} = 0,
\qquad 0 = 0 , \;\;\;
i \epsilon \; f_{2} + {\sqrt{\kappa} \over r}\; f_{1}  - m \; f_{5} = 0  ,
$$
$$
- i \epsilon \; f_{3} - {d \over dr} \; f_{1} - m \; f_{6} = 0 \qquad  0 = 0,
\qquad
- i ({d \over dr} + {1 \over r})\; f_{2} - i {\sqrt{\kappa} \over r} \; f_{3}
 - m \; f_{8} = 0 ,
$$
$$
 i {\sqrt{\kappa} \over r} \; f_{2}  - m \; f_{9}  =  0 ,\qquad    0 = 0 \; .
\eqno(2.10b)
$$

In an analogous way one can consider the $j = \mid \kappa \mid $   states
at negative $\kappa$:
\vspace{5mm}
$
\kappa  \leq  -1/2 :\qquad $
$$
\Psi = \exp{-i\epsilon t} \; [\; f_{1}\; D_{\kappa } ,\; 0 , \;f_{3} \;D_{\kappa},
\; f_{4} \;D_{\kappa +1} ,\; 0 ,
$$
$$
f_{6}\;  D_{\kappa } ,\; f_{7} \;D_{\kappa +1},
\; 0 , \;f_{9}\; D_{\kappa } ,\; f_{10} \; D_{\kappa +1}\; ]  \; ;
\eqno(2.11a)
$$
\vspace{3mm}
$$
( {d \over dr} + {2 \over r} ) \; f_{6} + {\sqrt{-\kappa} \over r}\; f_{7}
 + m \; f_{1}  = 0 \; , \;\;
0 = 0 \qquad  i \epsilon \; f_{6} - i {\sqrt{-\kappa} \over r} \; f_{10}
 - m \; f_{3} = 0  \; ,
$$
$$
i\epsilon \; f_{7} - i {\sqrt{-\kappa} \over r} \; f_{9} -
i({d \over dr} + {1 \over r}) \; f_{10} - m \; f_{4} = 0 \;\; , \qquad
0 = 0 \; ,
$$
$$
i \epsilon \; f_{3} + {d \over dr} \;  f_{1} +  m \; f_{6} = 0 ,
\;\; - i \epsilon \; f_{4} + {\sqrt{-\kappa} \over r} \; f_{1} + m \;f_{7} = 0 ,
\qquad  0 = 0  \; ,
$$
$$
 i {\sqrt{-\kappa} \over r} \; f_{4} + m \; f_{9}  = 0  , \qquad
i ({d \over dr} + {1 \over r}) \; f_{4} +  i {\sqrt{-\kappa} \over r} \;
 f_{3}  - m \; f_{10}  = 0   \; .
\eqno(2.11b)
$$

Thus,  the  task  of  finding  radial  equations   has   been
completely solved. All those systems look rather involved,  so  we
are reasons to question its easy analysis in terms of any standard
special  functions.  It  can  be  noted  that  the  ten  equations
established above fall naturally into 4 plus 6  sub-groups:  those
six give us a possibility to express the functions $f_{5},\ldots
,f_{10}$  in terms of $f_{1},\ldots, f_{4}$. Thereby, we can reduce
the~first order  system of $10$ equations to a~second order system  of
$4$  ones.  Evidently,
those four relation will represent a~still complicated system.

\subsection*{3. On connection with the Proca approach}

At  analyzing  the~above  radial  system,   any   additional
information can be useful. In particularly, as well  known,  there
must exist a first order  differential  condition  on  the  vector
constituent  of  $10$-dimensional  wave  function,   namely,
the~so-called generalized  Lorentz  relation.   Let  us  work  out  it
explicitly in this monopole situation.  To this  end,  instead  of
D-K formalism  it  will  be  more  convenient  to  use  the  Proca
formalism (see Sec.2):
$$
D_{\alpha } \; \Psi _{\beta } - D_{\beta } \; \Psi _{\alpha } =
{mc \over \hbar} \; \Psi _{\alpha \beta } ,\qquad
D^{\alpha } \;\Psi _{\alpha \beta } = {mc \over \hbar} \; \Psi _{\beta }
\eqno(3.1a)
$$

\noindent where $D_{\alpha }  = (\nabla _{\alpha } +
i\; {e \over \hbar c}\; A_{\alpha })$ ; $A_{\alpha }$  is  an~electromagnetic
potential (here,  it is presented by Scwinger  monopole  potential
$A_{\phi } = g\; \cos \phi )$. After the~operator $D_{\alpha }$  acts
on the~second equation in (3.1a), we will get
$$
{mc \over \hbar} \; ( \nabla _{\alpha } \; + \;
i \; {e \over c \hbar} \; A_{\alpha }) \; \Psi ^{\alpha } =
i \; {e \over 2c \hbar} \; F_{\alpha \beta } \; \Psi ^{\alpha \beta }
\eqno(3.1b)
$$

\noindent where, $F_{\alpha \beta } = (\partial _{\alpha } \; A_{\beta } -
\partial _{\beta } \; A_{\alpha } )$. When $A_{\alpha } = 0$, this relations
provides us with the usual Lorentz condition
$\nabla _{\alpha } \; \Psi ^{\alpha } = 0$.

Now, we face to  translate  this  relationship (3.1b) from
Proca representation into the Duffin-Kemmer's.  All above, instead
of $\Psi ^{\alpha }$  and $\Psi ^{\alpha \beta}$  we  have to introduce
 their tetrad components:
$
\Psi ^{\alpha } = e^{(a)\alpha } \; \Psi _{a} , \;\;\;
\Psi ^{\alpha \beta }  = e^{(a)\alpha } \; e^{(b)\beta } \; \Psi _{ab} .
$ Correspondingly, the (3.1b) will take on the~form
$$
{mc \over \hbar} \; [\; e^{(a)\alpha }_{;\alpha } \; \Psi _{a} \; +  \;
e^{(a)\alpha } \; \partial _{\alpha } \; ] \; \Psi _{a} \; + \;
 \; i\; {e \over \hbar c} \; A^{a}\; \Psi _{a} \; ] \; =
\; i\; {e \over 2\hbar c} \;
 F ^{ab} \; \Psi _{ab}
\eqno(3.1c)
$$

\noindent The coordinate representatives of the monopole
$ A_{\phi } = g\; \cos\theta  , \;\; F_{\theta \phi } = - g \sin \theta$
have the following tetrad description
$$
A^{2} = e^{(2) \phi }\; A_{\phi } = - g {\cos \theta  \over r \sin \theta } , \qquad
F^{12} \;= \; e^{(1)\theta } \; e^{(2)\phi } \; F_{\theta,\phi} = \; - \; {g \over r^{2}}
\eqno(3.1d)
$$

\noindent In addition, on simple straightforward computation, we find
$$
e^{(0)\alpha }_{\;\;\; ;\alpha } = 0 , \;\;
e^{(1)\alpha }_{\;\;\; ;\alpha } = - {\cos \theta \over r \sin \theta } ,\;\;
e^{(2)\alpha }_{\;\;\; ;\alpha } = 0 , \;\; e^{(3)A}_{\;\;\; ;\alpha } = - {2 \over  r}
\eqno(3.1e)
$$

\noindent The functions $\Psi _{a}$ and $\Psi _{ab}$ involved in (3.1c),
relate  to  the  $10$ constituents of $D-K$  column $\Phi$  as follows
 (this  represents  translating  from cyclic basis into Cartesian one;
$W \equiv -1/ \sqrt{2}$)
$$
\left ( \begin{array}{c}
\Phi_{0} \\ \Phi_{1} \\ \Phi_{2} \\ \Phi_{3} \\
         \Phi_{01} \\ \Phi_{02} \\ \Phi_{03} \\
         \Phi_{23} \\ \Phi_{31} \\ \Phi_{12}
\end{array} \right ) =
\left ( \begin{array}{cccccccccc}
1      & 0 & 0 &  0      & 0 & 0 & 0      & 0 & 0 &  0 \\
0      &-W & 0 & +W      & 0 & 0 & 0      & 0 & 0 &  0 \\
0      &-iW& 0 &-iW      & 0 & 0 & 0      & 0 & 0 &  0 \\
0      & 0 & 1 &  0      & 0 & 0 & 0      & 0 & 0 &  0 \\
0      & 0 & 0 & 0       &-W & 0 & +W     & 0 & 0 &  0 \\
0      & 0 & 0 & 0       &-iW& 0 &-iW     & 0 & 0 &  0 \\
0      & 0 & 0 & 0       & 0 & 1 & 0      & 0 & 0 &  0 \\
0      & 0 & 0 & 0       & 0 & 0 & 0      &-W & 0 & +W \\
0      & 0 & 0 & 0       & 0 & 0 & 0      &-iW& 0 &-iW \\
0      & 0 & 0 & 0       & 0 & 1 & 0      & 0 & 0 &  0
\end{array} \right ) =
\left ( \begin{array}{c}
    f_{1} \;D_{\kappa}    \\  f_{2} \;D_{\kappa-1} \\  f_{3} \;D_{\kappa} \\
    f_{4} \; D_{\kappa+1} \\  f_{5} \; D_{\kappa-1} \\  f_{6} D_{\kappa}   \\
    f_{7} \; D_{\kappa+1} \\  f_{8} \;D_{\kappa-1}  \\  f_{9} D_{\kappa}   \\
    f_{10}\;D_{\kappa+1}
\end{array} \right )
\eqno(3.2a)
$$

\noindent In the following we need only the components
$\Psi _{0}, \;\Psi _{1} , \;\Psi _{2} , \;\Psi _{3} , \;\Psi _{12}:$
$$
\Psi _{0} = \; e^{-i\epsilon t} \; f_{1} \; D_{\kappa } , \qquad
\Psi _{3} = \; e^{-i\epsilon t} \; f_{3} \; D_{\kappa } , \qquad
\Psi _{1} = \; e^{-i\epsilon t} {1 \over \sqrt{2}} \;
(- f_{2} \;D_{\kappa -1} + f_{4} \; D_{\kappa +1} ) ,
$$
$$
\Psi _{2} = \; e^{-i\epsilon t} \; {i \over \sqrt{2}} \;
(- f_{2} \;D_{\kappa -1}   - f_{4} \; D_{\kappa +1} ) , \qquad
\Psi _{12} = e^{-i\epsilon t} \; f_{9}\; D_{\kappa }
\eqno(3.2b)
$$

\noindent Allowing for (3.2b) and (3.1d,e),  the~condition (3.1c)
has taken the form:
$$
{mc \over \hbar} \; [\; {1 \over \sqrt{2}} \;  r \; f_{2} \;
( \partial _{\theta } \; D_{\kappa -1}  -
{m + (\kappa -1) \cos \theta \over \sin \theta} \;  D_{\kappa -1} )  -
{1 \over \sqrt{2}} \; r \;  f_{4}\;
( \partial _{\theta } \; D_{\kappa +1} +
$$
$$
{m + (\kappa +1) \cos \theta ) \over \sin \theta} \;  D _{\kappa +1}) + \;
 D_{\kappa } \;  ( - {2 \over r} \; f_{3} \; - \;
i {\epsilon \over \hbar c} \; f_{1} \; - \; {d \over dr}\; f_{3} ) ] =
 - i \; {\kappa \over r^{2}} \; f_{9} D_{\kappa}
\eqno(3.3a)
$$

\noindent After having used the recursive relations
$$
\partial _{\theta } \; D_{\kappa -1} -
{m + (\kappa -1)\cos \theta \over \sin \theta } \; D_{\kappa -1} =
- \sqrt{(j-\kappa +1)(j+\kappa )} \; D_{\kappa } ,
$$
$$
\partial _{\theta } \; D_{\kappa +1} -
{m + (\kappa +1)\cos \theta \over \sin \theta } \; D_{\kappa +1} =
- \sqrt{(j+\kappa +1)(j-\kappa )} \; D_{\kappa }
$$

\noindent (which are easily derived  from  the  used  above)  we  eventually
arrive at
$$
{mc \over \hbar} \;
- i \; {\epsilon \over \hbar c} \; f_{1} \;  + \; ({d \over dr} + {2 \over r})\;
f_{3}\; - \; {1 \over \sqrt{2}} \; ( c \; f_{2} + d \; f_{4})  =
 - i \; {\kappa \over r^{2}} \;  f_{9}
\eqno(3.3b)
$$

\noindent If $j = \mid \kappa \mid  , \kappa  \geq  +1/2$ , one gets
$$
{mc \over \hbar} \; [- i{\epsilon \over \hbar c} \; f_{1}  \; +  \;
 ( {d \over dr} + {2 \over r} ) \; f_{3}  \; -  \;
{\sqrt{\kappa} \over r} \; f_{2} ] = - i \; { \kappa \over r^{2}} \; f_{9}
\eqno(3.3c)
$$

\noindent if $j = \mid \kappa \mid , \kappa  \leq -1/2$,  one gets
$$
{mc \over \hbar} \; [- i \; {\epsilon \over \hbar c } \; f_{1} \; + \;
 ( {d \over dr} \; + \; {2 \over r} ) \; f_{3} -
{\sqrt{-\kappa} \over r} \; f_{2} \;] = - i \; { \kappa \over r^{2}} \; f_{9} \; .
\eqno(3.3d)
$$

\subsection*{4. Discret symmetry.}

Now,  let  us   take   up   else   one    question,   namely,
concerning a problem of discrete symmetry at the vector particle -
monopole case. As was shown in [36], at  the  electron-monopole
case there exists some composite operator
 $\hat{N} = [ \hat{\pi } \otimes \hat{P}_{bisp.} \otimes  \hat{P} ) ]$.
It would seems  that  the~same possibility is realized  also in  case of
vector   particle.
Indeed, a~direct extension  of  the~above  to  a~new  situation:
$\hat{N}_{vect.} = [ \hat{\pi } \otimes \hat{P}_{vect.} \otimes  \hat{P} )]$
affords formally an operator  with
analogous commuting properties, that is,
$$
[ \hat{N}_{vect.}, \hat{H}_{vect.} ]_{-} = 0 \; , \qquad
[ \hat{N}_{vect.}, \vec{J}^{eg}_{vect.} ]_{-} = 0   \; .
$$

\noindent However, as it will be verified bellow, such an operator cannot be
diagonalized on Duffin-Kemmer wave  functions  found  above.  This
matter is worth considering in more detail.

The  vector  ordinary $P$-reflection  operator  in  Cartesian
tetrad, is
$$
\hat{P}_{Cart.} \; = \;
\left ( \begin{array}{cccc}
     1  &   0  &  0  &  0  \\
     0  &  -I  &  0  &  0  \\
     0  &   0  & -I  &  0  \\
     0  &   0  &  0  & +I
\end{array} \right )
\eqno(4.1)
$$

\noindent where  a symbol "I" denotes a~unit $3\times 3$ matrix.
After translating this $\hat{P}_{Cart.}$ -operator into the~spherical
tetrad's  basis  according  to
$ \hat{P}_{sph.} = O(\theta ,\phi ) \;  \hat{P}_{Cart.}\;
 O^{-1}(\theta ,\phi )$,
where $( O(\theta ,\phi)$  is  a  $10$-dimension
rotational  matrix associated with the~spinor  gauge  transformation used
in case of electronic field, it takes on the~form (the standart cyclic basis
in the vrctor space is used)
$$
\hat{P}_{sph.}^{cycl.} \; = \;
\left ( \begin{array}{cccc}
     1  &   0  &  0  &  0  \\
     0  &  +E  &  0  &  0  \\
     0  &   0  & +E  &  0  \\
     0  &   0  &  0  & -E
\end{array} \right ) , \qquad \hbox{where} \qquad
E \equiv \left ( \begin{array}{ccc}
          0   &  0  & 1 \\ 0  & 1  &  0 \\  1  &  0  &  0
\end{array} \right )    \; .
\eqno(4.2)
$$
\noindent From the equation on proper values
$$
[ \hat{\pi } \otimes \hat{P}^{cycl.}_{sph.} \otimes \hat{P} ] \;
 \Phi ^{eg}_{jm} = N \; \Phi ^{eg}_{jm}
\eqno(4.3a)
$$

\noindent it follows
$$
N = (-1)^{j+1} : \qquad  f_{1} = \; f_{3} = \; f_{6} = \; 0 ,
\; f_{4} = - f_{2}, \; f_{7} = - f_{5} , \;  f_{10} = + f_{8} ;
\eqno(4.3b)
$$
$$
N = (-1)^{j} : \qquad f_{9} = 0 , \;
f_{4} = + \; f_{2} , \;  f_{7}  = +\;  f_{5} ,\;  f_{10} = -\; f_{8}
\eqno(4.3c)
$$

\noindent these  relations  are  exactly   the   same   which  had  arisen  from
diagonalizing the~ordinary $P$-reflection operator  in  case  of
a~free vector field:
$[ \hat{P}^{cycl.}_{sph.} \otimes \hat{P} ] \Phi^{0} = P \Phi^{0}$.

Let us try imposing these  additional  relations (4.3b)  or
(4.3c) on radial functions $f_{1}(r), \ldots
, f_{10}(r)$ obeying the system
(2.4). On direct verification , one concludes  that  a~system  so
achieved is not self-consistent. This means that the $\hat{N}$  operator,
though  commuting   with  the  vector $eg$-Hamiltonian,  cannot  be
regarded as an observable quantity  measured  simultaneously  with
vector particle-monopole's  Hamiltonian.  For  example,  in  case
(4.3b), one has
$$
- ( {d \over dr} + {2 \over r} )\; 0 - {\sqrt{2} \over r}\;
 ( c - d )\; f_{5} - m\; 0  =  0  \; ,
\qquad \hbox{that is} \qquad  f_{5} \equiv  0  \;  ;
$$
$$
i \epsilon \; 0 + i ({d \over dr}  + {1 \over r} ) \; f_{8} +
i{\sqrt{2} c \over r} \;  f_{9}  - m  \; f_{2} = 0  \; ;
$$
$$
i \epsilon \;  0 + 2 {i \over r} \; (- c + d ) \;  f_{8} - m \; 0 = 0 \; ,
\qquad \hbox{that is} \qquad  f_{8} \equiv  0  \;
$$
$$
i \epsilon \; 0 - i ( {d \over dr} +  {1 \over r} ) \; 0 -
{\sqrt{2} d \over r}\;  f_{9} - m \;  f_{2} = 0  \; ;
$$
$$
i \epsilon  f_{2} + {\sqrt{2} c \over r} \; 0  - m \; 0 = 0   \; ,
\qquad \hbox{that is} \qquad  f_{2} \equiv  0 \; , \; f_{9} \equiv 0 \; ,
$$
$$
- i \epsilon \;  0 - {d \over dr}  0 - m \; 0 = 0  \;\; \; , \;\;
- i \epsilon \; 0 + {\sqrt{2} d \over r})\; 0 - m \; ,
$$
$$
0 = 0 \;\; ,  \;\;
- i ({d \over dr} + {1 \over r}) \; 0  - i{\sqrt{2} c \over r} \; 0
 - m\; 0 = 0                       \; .
$$

\noindent So, all the $f_{i}(r)$ turn out  to  be  equal  to  zero;
but  such  a~solution is not of interest because of its triviality.

Here one gives some added comment  on  extending
the~vector particle-monopole  formalism  constructed  above  to
an arbitrary background space-time with  spherical  symmetry.  The
relevant Duffin-Kemmer $eg$-equation is taken in the form
$$
\left [ i \beta  ^{0} \; (e^{-\nu /2} \; \partial  _{t} \; + \;
{1 \over 2}\; {\partial \nu \over \partial r} \; e ^{-\mu /2}\; j ^{03}) \; + \;
+ i \beta ^{3} \; (e^{-\mu /2} \; \partial _{r}  \; +  \;
{1 \over 2} \; {\partial \mu \over \partial t} \;
e^{-\nu /2} \; j^{03} ) \; +     \right.
$$
$$
\left.
- \; {i \over r}\; e^{-\mu /2} \; ( \beta ^{1} \; \beta ^{12} \; + \;
\beta ^{2} \; \beta ^{23}) \; + \;
{1 \over r} \; \Sigma _{\theta ,\phi } \; -
\; {mc \over \hbar} \;  \right ] \; \Phi (x) = 0 \; .
$$

\noindent Therefore, almost all done above for
the flat space model will be easily  taken  into  a~curved  space
model with only several evident changes.

\end{document}